\documentclass[sigconf,nonacm,11pt]{acmart}
\makeatletter
\def\@ACM@checkaffil{
    \if@ACM@instpresent\else
    \ClassWarningNoLine{\@classname}{No institution present for an affiliation}%
    \fi
    \if@ACM@citypresent\else
    \ClassWarningNoLine{\@classname}{No city present for an affiliation}%
    \fi
    \if@ACM@countrypresent\else
        \ClassWarningNoLine{\@classname}{No country present for an affiliation}%
    \fi
}
\makeatother

\AtBeginDocument{%
  \providecommand\BibTeX{{%
    \normalfont B\kern-0.5em{\scshape i\kern-0.25em b}\kern-0.8em\TeX}}}

\graphicspath{{fig/}{./}}

\copyrightyear{2019}
\acmYear{2019}
\setcopyright{rightsretained}

\acmConference{CSE6250}
\acmDOI{Big Data For Health}
\acmISBN{}
\acmBooktitle{}

\begin{document}

\title{ LeDNet: Localization-enabled Deep Neural Network for Multi-Label Radiography Image Classification}


\author{Lalit Pant}
\email{lpant3@gatech.edu}
\affiliation{%
 }

\author{Shubham Arora}
\email{shubham.arora@gatech.edu}
\affiliation{%
}




\begin{abstract}
\textbf{Multi-label radiography image classification has long been a topic of interest in neural networks research. In this paper, we intend to classify such images using convolution neural networks with novel localization techniques. We will use the chest x-ray images to detect thoracic diseases for this purpose. For accurate diagnosis, it is crucial to train the network with good quality images. But many chest X-ray images have irrelevant external objects like distractions created by faulty scans, electronic devices scanned next to lung region, scans inadvertently capturing bodily air etc. To address these, we propose a combination of localization and deep learning algorithms called LeDNet to predict thoracic diseases with higher accuracy. We identify and extract the lung region masks from chest x-ray images through localization. These  masks are superimposed on the original X-ray images to create the mask overlay images. DenseNet-121 classification models are then used for feature selection to retrieve features of the entire chest X-ray images and the localized mask overlay images. These features are then used to predict disease classification. Our experiments involve comparing classification results obtained with original CheXpert images and mask overlay images. The comparison is demonstrated through accuracy and loss curve analyses.} 
\end{abstract}


\maketitle
\section{Introduction}
Multi-label radiography images can give highly inaccurate results with object classification algorithms due to many irrelevant objects present in them. Such images are specially prevalent in the field of chest radiology imaging. Chest radiography is the most common imaging technique which helps identify many life-threatening thoracic diseases by examining the lung regions. With the availability of a large-scale data set \cite{xraydatabase}, several studies have been conducted to automatically detect abnormalities in chest X-rays. The work published by Wang et al. \cite{xraydatabase} evaluated four Convolutional Neural Networks (CNN) architectures (AlexNet, VGGNet, GoogLeNet and ResNet \cite{residual}). Rajpurkar et al. \cite{radiologist} demonstrated a DenseNet architecture that provides effective performance for thoracic disease classification. A reference implementation for DenseNet using PyTorch is provided in CheXNet \cite{pytorch}.
Following attempts to process region-based Convolutional Neural Networks \cite{ieee} \cite{pyramid}, Liu et al. \cite{segmentation} present a novel approach termed as SDFN, which crops local lungs regions using higher-resolution information from JSRT database \cite{chestradiograph}. Along with demonstrating more accurate disease classification, the experiments proved to localize the lesion regions more precisely as compared to the traditional method. Zongyuan Ge et al. \cite{multilabel} propose a deep network architecture implementing an error function called multi-label softmax loss (MSML) that handles presence of multiple data labels and imbalanced classes which is common in medical problems. Tao \cite{deep} gives us the intuition about the role of Ensemble Deep Neural Network to help prevent the overfitting problem and how this can be leveraged for imbalanced data.

\section{Problem Formulation}

Chest X-ray images are typically affected by data quality issues like faulty scans, other electronic devices captured in the images or interference due to bodily air. All these factors can lead to false positive identification and incorrect prediction outcomes. To be most effective in predicting the thoracic diseases, the approaches discussed above would need clean and localized images to be fed to the deep learning network. After understanding the CheXpert \cite{chexpert} data which is multi-label and has imbalanced classes, we believe that applying localization on lung region followed by deep learning with DenseNet can greatly improve the performance of these state-of-arts models for thoracic disease classification.

This leads us to present a deep learning architecture in this paper termed as Localization-enabled Deep Neural Network (LeDNet), which leverages the localized lung region images. The underlying intuition behind this study is that the thoracic diseases are mostly restricted to the lung regions. If the lung regions are identified, the networks can be trained using lung images without irrelevant regions. Additionally, if the disease is associated with other features outside of the lung regions, the entire chest X-ray images can also be considered. A detailed description of the experimental methodology, the data used, the results and future optimizations are provided in this study.

\section{Methodology}
Two publicly available datasets are used in this study.
The first dataset is the JSRT Dataset \cite{chestradiograph} which contains 154 nodule and 93 non-nodule chest X-ray images. Each image has a resolution of 2048X2048 pixels with a 12-bit density range. The masks for the lung region are also publicly available at SCR database \cite{chestimages}. This dataset has been used to train the localization model which helps us remove irregularities around lung region.

The second dataset is CheXpert (Chest eXpert), \cite{chexpert} a large dataset for chest radiograph interpretation. The dataset consists of 224,316 chest radiographs (consisting of frontal and lateral views both) of 65,240 patients labeled for the presence of 14 common chest radiographic observations which can be positive, negative and uncertain. We are processing this dataset through localization-trained model to get improved lung images. We use only the frontal chest X-ray images of this dataset because the lateral images could be wrongly localized.

The proposed LeDNet is composed of an analytics pipeline with below modules:

\textbf{a. Pre-processing and Localization Training}: Using the JSRT dataset, we train an image localization model which automatically extracts the lung-region localized images from given chest X-ray images. This uses the UNet implementation as proposed in \cite{lung}. The idea behind localization is that the thoracic diseases are generally limited to the lung regions.
If lung-region localized images can be produced, then the networks can be trained without noisy regions in data, leading to improved results by the predictive model. Original JST images are pre-processed with histogram localization while the left and right mask images provided by SCR database are combined into a single image as depicted in Figure 1.
  \begin{figure}[h]
  \centering
  \includegraphics[width=\linewidth]{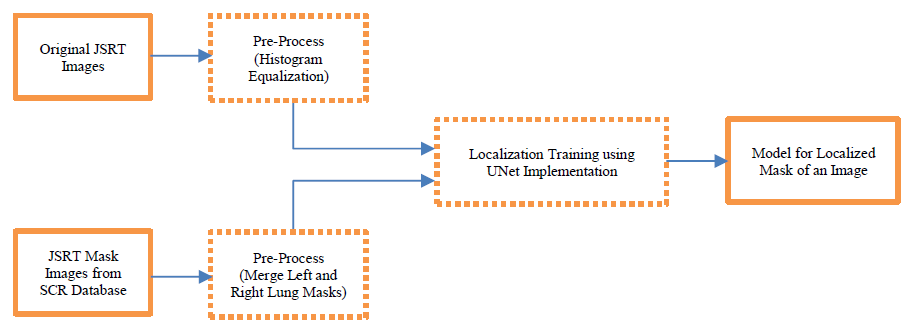}
  \caption{The Pre-process and Localization Training module}
  \end{figure}

\textbf{b. Localization Prediction and Mask Overlay}:  This pipeline runs localization/mask prediction on the original images. We then overlay the localized(masked) images on top of the original CheXpert images to extract lung image portions from it. Extracting from original images gives us an ability to train the model on actual lung regions (with 3 RGB channel) rather than localized binary images, producing improved results. Later on, we will compare the disease classification results from overlay images with the one from original CheXpert images.
  \begin{figure}[h]
  \centering
  \includegraphics[width=\linewidth]{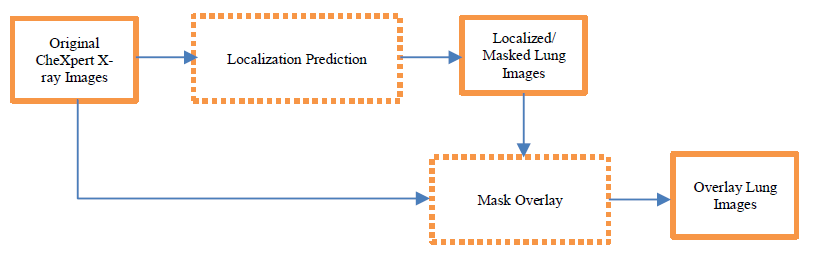}
  \caption{The Localization Prediction and Overlay module}
  \end{figure}
  
\textbf{c. CheXpert Pre-processing}:
The given CheXpert dataset has large number of uncertain labels, empty or negative value in loss can give Nan loss function which results in failure while training the algorithm. In this implementation we have considered such labels to be 0 i.e. disease not identified. While reading the data we have also performed normalization of images to make them of same size and reduce abnormalities.
  \begin{figure}[h]
  \centering
  \includegraphics[width=\linewidth]{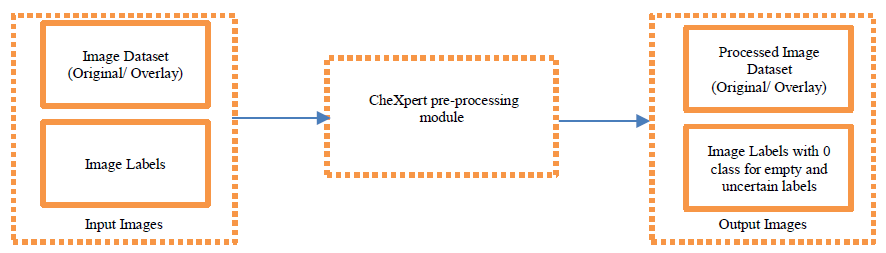}
  \caption{The pre-processing module to classify empty and uncertain labels}
  \end{figure}

\textbf{d. Feature Selection and Prediction}:
The DenseNet-121 has been reported to be giving state-of-the arts performance for thoracic disease classification [9]. So, the output of localized images is passed through it. Since the localized images highlight the important regions of the image, it is expected that the model will give better outcomes. While training the algorithm we tried multiple other algorithms like Alexnet and ResNet-50, even though AlexNet and ResNet-50 gave the outcomes faster than DenseNet-121. We choose DenseNet-121 because of the accuracy.

CheXpert dataset is a multilabel classification problem, while reading these images we tried two approaches of classification i.e. predicting same image with multiple classification Y i.e. {(X,Y1), (X,Y2), ... (X,Yn)} and Y as one vector i.e. {X,(Y1,Y2,...Yn)}. For us second approach gave better results and performance. Dataset also required preprocessing around the labels as many of the images either marked as uncertain or with no (NaN) label. So, we pre-processed such labels and considered them as 0 or no disease.

Being a multilabel classification problem we utilized Binary Cross Entropy loss function with Adam Optimizer. The Dense-Net121 model is followed by a sigmoid function to normalize the output. Since the outcome of sigmoid function is always expected to be between 0 and 1, we calculated accuracy by rounding off the output and comparing it with target. This helped us label predicted output to binary classification. We also weighed probability of loss function tending to NaN value, but since the values are expected to be between 0 and 1 we didn’t utilize techniques like softmax or normalization over the outcome.

For better comparison this pipeline was triggered with localized and original CheXpert images and we found localized images gave better outcome.
  \begin{figure}[h]
  \centering
  \includegraphics[width=\linewidth]{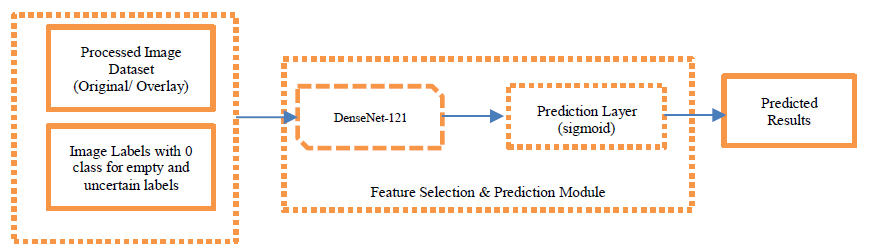}
  \caption{The Feature Selection and Prediction module}
  \end{figure}

We currently use accuracy and loss curves as a comparison metrics for the proposed DenseNet-121 model. We use IoU (also called Jaccard Index) and Dice Coefficient to evaluate the training of localization model as these are standard metrics for such type of segmentation related models. We also use accuracy and loss values for evaluating the localization model.

\section{Experimental Results}
This analytics pipeline is setup on Azure NC6 GPU machine to support image processing for the small CheXpert dataset (approx. 11G). For localization module have utilized JSRT Dataset of 247 images for giving the trained model. This trained model was utilized with CheXpert dataset to provide the localized dataset. Classification was on 28629 subset images CheXpert dataset. Python is our programming language of choice for all modules and we utilized using python libraries like pandas, numpy and scipy for data analysis. For DenseNet implementation, we are using PyTorch along with Torch Vision API for image processing. For localization, we are using existing UNet implementation from UNet repository \cite{lung} which use Keras.

\textbf{a. Localization}: 
We ran the localization model training with 8 batch size and 10 epochs after splitting 247 JSRT images into training (80\%) and validation (20\%) sets. Training the UNet implementation [12] of the localization model on complete JSRT dataset and its mask images provided below results which can be considered highly accurate:

Mean IoU: 0.9613

Mean Dice: 0.9803

Also, below are some other evaluation metrics which show good performance on both training and validation sets:

Train loss: 0.2369 – Train Accuracy: 0.9083

Valid Loss: 0.1626 – Valid Accuracy: 0.9424

We found that the pre-trained model available from U-Net implementation has Mean IoU of 0.9710 and Mean Dice of 0.9850 on the JSRT dataset which mean pre-trained model can perform better than the above trained model. Hence, we concluded to use the pre-trained model for localization prediction on CheXpert dataset.

With limited GPU resources and multiple deep learning models to execute (related to localization and classification both), we decided to run the pre-trained localization model on only a subset of CheXpert images (28,629 to be exact). This decision helped us to limit the scope and focus on completing each pipeline with available compute resources.

\textbf{b. Classification}:
28,629 images of CheXpert dataset is divided into train, validate and test sub datasets to evaluate the model with 70, 20 and 10 ratios respectively. CheXpert dataset is also pre-processed to set uncertain and empty labels as 0. Input images can be of different sizes, so we have performed transformations over the dataset to resize the images to 256 X 256 with random horizontal flips to allow the network to be trained with different varieties of data.
We ran the explained pipeline with original and overlay images to get the outcomes as in Figure 5.
  \begin{figure}[h]
  \centering
  \includegraphics[width=\linewidth]{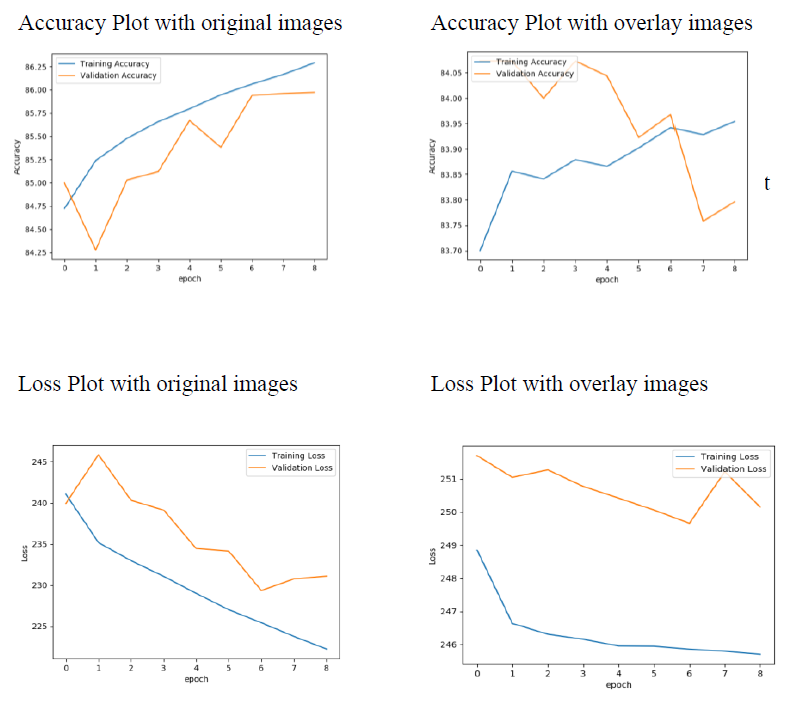}
  \caption{Accuracy Plots and Loss Plots for DenseNet-121 on CheXpert original and overlay images}
  \end{figure}
\section{Experimental Evaluation}
In this study, we are developing a novel implementation called LeDNet for thoracic disease classification on CheXpert dataset using localization model derived from JSRT dataset which helps us identify lung regions and in turn, improves classification performance. The primary intuition of this study is the emphasis on the most meaningful part of the CheXpert images (i.e. lung regions) and compare the performance improvement using it.

CheXpert dataset has variety of images with lots of irregularities. One such irregularity is unclassified images, which comprises of a big chunk of the dataset, thus dropping those images from training dataset may not give us great outcome and also result in NaN loss values. In this study, we classified those datasets as 0 i.e. disease not identified. This may result in wrong interpretations of data and will require future evaluation to try techniques like self-training.

While analyzing the outcomes, we came to an interesting conclusion that if the images are localized, we may lose important features which are important for the label’s prediction. To get the best of both the models we decided to overlay the localized images over the original dataset so that we don’t miss on important features. Future work will involve combining features of localized and original images at the end of the network. First model did not provide better results as expected and resulted in degradation of accuracy with the increase in number of epochs which is showcased in above graphs. We expect our second approach merging of outcomes of localized and original images could result in better outcomes.

Another important observation was that if we run all the pipelines together end-to-end, the network becomes compute intensive and has runtime performance issues. In order to mitigate this problem, we split our approach into 2 individual pipelines with pre-processing steps in both which we ran independently in separate networks (i.e. as separate processes).

While running classification module, we ran into multiple issues due to limited compute resources and ended with 50 batch size for running the job. We started our model training and validation with smaller number of epochs but after analyzing the accuracy and loss function statistics we were not able to figure out the trend around prediction statistics. It was after 4-5 epochs when the overlay model started predicting with less accuracy and graph started following the downward path. Thus, we tried our training with maximum 8 epochs to deliver better outcome. Figure 6 and 7 shows the statistics at different epochs.
  \begin{figure}[h]
  \centering
  \includegraphics[width=\linewidth]{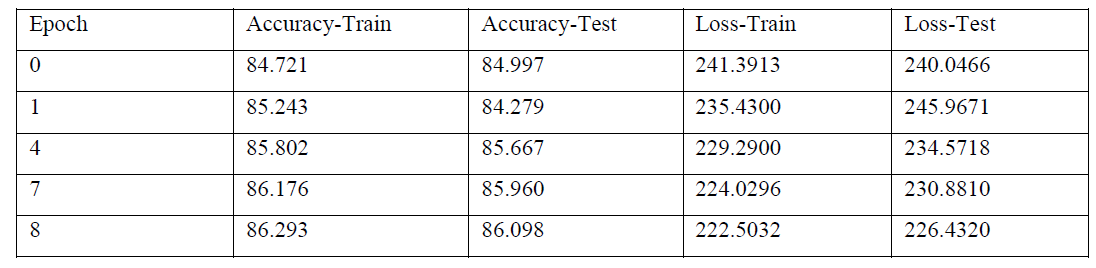}
  \caption{Table for CheXpert dataset}
  \end{figure}

    \begin{figure}[h]
  \centering
  \includegraphics[width=\linewidth]{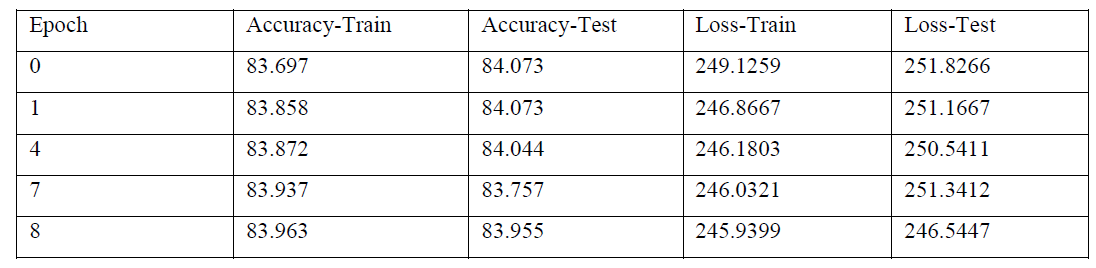}
  \caption{Table for Overlay dataset}
  \end{figure}
From these results it is evident that with original dataset accuracy improved over the time and on the other hand with overlay/localized dataset model accuracy and loss function value remain steady across epochs.
 
\section{Future Work}

In some cases, we observed that the localized CheXpert lung images had more than two regions which leads to false positive identification. This can primarily be attributed to external air in the original images and thus located further away from the image center than the lungs. To mitigate this issue, we can look to post-process the localized CheXpert images to retain only the two regions with shortest distance from image center. Also, in cases where only 1 localized region is located in the images, we can obtain the other region by mirroring the located region along to vertical centerline of image. Both these post-processing steps improved the localized image quality and have the potential to improve model performance significantly.

Uncertainty labels are quite prevalent in this dataset for some observations for example, 12.78\% in comparison to 6.78\% of positives for consolidations. We can use self-training for utilizing uncertain labels during model training. This can be accomplished by training a self-training model considering uncertain labels as unlabeled data, and then
using the model to re-label the uncertain labels with predicted value given by the model until convergence. This self-training approach defined in \cite{chexpert} which can help us utilize this dataset in its entirety.

The CheXpert dataset is highly imbalanced with more normal cases than abnormalities, so we can use Ensemble DenseNet as it avoids overfitting and can give improved results with imbalanced data.

Original chest X-ray images and localized images can both contain vital characteristics for disease classification, hence we can take full advantage of both type of images by combining their generated features. One way of achieving it is by concatenating the Feature Selection output (DenseNet-121) from each image type and connecting it to a 14-Dimension Fully Connected prediction layer with sigmoid activation function.

\section{Conclusion}

Prediction of diseases with Chest X-ray images is novel problem and could be predicted with decent accuracy utilizing LeDNet. In this paper we detailed an approach of localization combined with deep neural network to predict diseases. We also compared outcomes with localization and without localization. This study opens gates for further evaluation of this approach with certain modifications related to localization and can provide better outcomes.

We have observed DenseNet-121 training time to be compute intensive, hence considering the size of CheXpert dataset, running this experiment on Azure with GPUs helped us performing the predictions faster.

In future we will like to run the pipeline with CheXpert original dataset and verify if the outcomes could be better. We would also like to perform self-training for utilizing the uncertain labels in CheXpert dataset and lastly, utilize ensemble bagging technique to reduce overfitting.

\end{document}